\documentclass[journal=jacsat,manuscript=article]{achemso}
\SectionNumbersOn
\usepackage{hyperref}
\usepackage{natbib}
\usepackage[version=3]{mhchem} % Formula subscripts using \ce{}
\usepackage{graphicx}
\usepackage{amsmath}
\usepackage{amssymb}
\usepackage{bm}
\usepackage{latexsym}
\usepackage{epsfig,epsf,rotating}
\usepackage{indentfirst}
\usepackage{epstopdf}
\usepackage{xcolor}
\usepackage{multirow}
\usepackage{tabularx}
\usepackage{ctable}
\usepackage{upgreek}
\usepackage{nicefrac}
\usepackage[utf8]{inputenc}
\usepackage[T1]{fontenc}
\usepackage{mathptmx}
\usepackage{etoolbox}
\usepackage{array}
\usepackage{pdfpages}

\makeatletter
    \newenvironment{codeavailability}{%
        \acs@section*{\codeavailabilityname}%
    }{}
    \newcommand*\codeavailabilityname{Code Availability}
    \makeatother

\author{Rohan S. Adhikari}
\email{rohana@udel.edu}
\affiliation[RiceChBE]
{Department of Chemical and Biomolecular Engineering, Rice University, 6100 Main St., Houston, TX 77005, USA}
\alsoaffiliation[UDCBE]
{Department of Chemical and Biomolecular Engineering, University of Delaware, Newark, DE 19716, USA}
\author{Winnie H. Shi}
\affiliation[RiceChBE]
{Department of Chemical and Biomolecular Engineering, Rice University, 6100 Main St., Houston, TX 77005, USA}
\author{Amanda B. Marciel}
\email{am152@rice.edu}
\affiliation[RiceChBE]
{Department of Chemical and Biomolecular Engineering, Rice University, 6100 Main St., Houston, TX 77005, USA}
\author{Walter G. Chapman}
\email{wgchap@rice.edu}
\affiliation[RiceChBE]
{Department of Chemical and Biomolecular Engineering, Rice University, 6100 Main St., Houston, TX 77005, USA}

\title[ACS AMBER FFs]
  {The Protein Force Field Plays a Crucial Role in Obtaining Accurate Macromolecular Ensembles of IDPs}

\abbreviations{SAXS, CMAP, R$_\mathrm{g}$}

\begin{document}


\clearpage

\begin{abstract}
  Intrinsically disordered proteins (IDPs) play a significant role in intracellular phenomena and are known to exist in an ensemble of inter-converting conformations in solution. Accurately modeling the conformations of IDPs in solution poses a challenge to traditional force fields that are tuned to predict the properties of folded proteins. There is a need for generalized atomistic force fields that can accurately predict the properties of both folded proteins and IDPs. Improvements to protein force fields for increased accuracy in secondary structure prediction and new water models with increased water-water dispersion interactions have been proposed in search of a generalized simulation method. Validating the proposed improvements against experiments poses challenges such as a lack of suitable systems to test the generalizability and choosing a property of interest to match the simulation results against experiments. In this work, we use small angle X-ray scattering (SAXS) data from peptide-based polyampholytes that mimic IDPs to test the generalizability of the AMBER protein force fields and the OPC water model. The specific improvements due to the AMBER ff19SB protein force field and the OPC water model are isolated and studied. Analysis of SAXS profiles and the conformational distribution of polyampholyte sequences show the AMBER ff19SB-OPC water combination to be a generalized model that predicts both ordered polyampholyte sequences and disordered polyampholyte sequences in good agreement with experiments. We have developed a new scattering model termed `SWAXS-AMDE' that accounts for the hydration layer density changes in atomic detail and is particularly useful in making one-to-one comparisons of simulated scattering profiles to experiments. SWAXS-AMDE allows for the thermal fluctuations of the solute which is particularly consequential for IDPs.
\end{abstract}

\section{Introduction}

A detailed analysis of protein structures deposited in the `SWISSPROT' database revealed 1000s of proteins that were at least partially disordered \cite{romero1998thousands, wootton1994non}. This class of proteins that exist in disordered conformations in solution have since come to be known as intrinsically disordered proteins (IDPs) \cite{tompa2012intrinsically, wright2015intrinsically}. IDPs are now recognized to play a vital role in membraneless organelle formation \cite{wei2017phase}, regulation of signaling networks \cite{bondos2022intrinsically, dyson2005intrinsically}, and other intracellular phenomena. These findings at least partially explain the abundance of IDPs in complex organisms and seriously challenge the long-held theory that proposed an innate connection between the structure of a protein in solution and its functionality (structure-function hypothesis) \cite{wright1999intrinsically}. Interestingly, perturbations that disrupt the endogenous condensate behavior of IDPs result in pathological states that have been implicated as a leading cause of a number of neurodegenerative disorders \cite{uversky2008amyloidogenesis, kulkarni2019intrinsically}. 

Molecular dynamics (MD) simulations can serve as a valuable tool in providing atomically detailed insights about the inter-conversion between IDP conformations \cite{chong2017computer, huang2018force}. Traditional force fields, however, were tuned to predict the conformations of folded proteins and have been shown to predict overly-compact conformations for IDPs \cite{rauscher2015structural, henriques2015molecular, huang2018force}. To circumvent the generation of over-compact IDP ensembles, two distinct philosophies of force field tuning have been explored recently. First, the dihedral angle correction maps (CMAPs) of protein force fields have been reweighted with the goal of accurately capturing the frequent $\mathrm{\beta}$-sheet to $\mathrm{\alpha}$-helix transitions in IDPs \cite{rauscher2015structural, huang2017charmm36m, huang2018force}. Second, new water models (e.g., TIP4P-D \cite{piana2015water} and OPC \cite{izadi2014building}) have been proposed with increased water-water dispersion interactions. Making the water molecules more cohesive by increasing the water-water dispersion interactions is also hypothesized to remove the over-compaction of IDP ensembles \cite{onufriev2018water}. The OPC water model was shown to generate extended conformations for representative IDPs in good agreement with the experimental radius of gyration (R$_\mathrm{g}$) estimates \cite{shabane2019general}. The mean R$_\mathrm{g}$ does not fully capture the properties of the ensemble of conformations generated by the force field. Validating the solution ensembles generated by force fields requires a more detailed comparison to experiments and suitable test systems for such a comparison. In this context, a class of materials called polyampholytes is extremely relevant. 

Polyampholytes are charged polymers that contain both cationic and anionic groups \cite{dobrynin2004polyampholytes, dinic2021polyampholyte}. Approximately 75\% of all known IDPs are polyampholytes \cite{kudaibergenov2021synthetic, romero2001sequence}. Specifically, the bioinspired EK polyampholytes that are made up of glutamic acid (E) and lysine (K) amino acid residues have garnered significant attention \cite{shi2023influence, mccarty2019complete, devarajan2022effect}. The low sequence complexity of EK polyampholytes and the propensity of the E and K amino acid residues to induce disorder make them idealized mimics of IDPs \cite{uversky2013alphabet, hansen2006intrinsic}. Moreover, the amino acid residues can be rearranged to obtain different sequences of polyampholytes. Implicit solvent studies on EK polyampholytes have suggested a transition from disordered conformations to stable preferred conformations in solution as the amino acid residues are rearranged from an alternating sequence to a diblock sequence \cite{das2013conformations}. For these reasons, the EK polyampholytes appear to be ideal test systems to validate the generalizability of atomistic force fields. To ensure a direct comparison between force field predictions and experimental data for polyampholytes, the choice of the experimental characterization technique should be made carefully.

In comparison to nuclear magnetic resonance (NMR) and F{\"o}rster resonance energy transfer (FRET), small angle X-ray scattering (SAXS) imposes fewer restrictions on sample preparation \cite{svergun2013small}. In order to infer the conformational properties of biomolecules, the scattering intensities of the buffer solvent (background) are subtracted from the scattering intensities of the biomolecular solution in a procedure called background subtraction \cite{lipfert2007small}. Computational methods that seek to compare background subtracted scattering intensities to experiments need to account for all electron density changes (contrast) that arise in the biomolecular solution relative to the bulk solvent. Contrasts arise both in the solvent regions with non-bulk like behavior and in the regions occupied by the solute. Regions with non-bulk like solvent densities can be divided into two categories; a region with excess solvent density in the layer immediately surrounding the solute, and a region with depleted solvent density in the volume occupied by the solute (excluded volume) \cite{svergun2013small}. Svergun et al. proposed the CRYSOL method which performs the scattering calculation by assuming a continuum density for the excess solvent in the layer immediately surrounding the solute and a continuum bulk density of the solvent in the excluded volume \cite{svergun1995crysol}. Park et al. proposed an explicit water scattering model (EWSM) which computes the scattering intensities by accounting for the solvent density changes in atomic detail \cite{park2009simulated}. Both CRYSOL and EWSM assume a frozen conformation for the solute. The WAXSiS \cite{knight2015waxsis} program proposed by Knight and Hub calculates the scattering profile by accounting for the solvent density changes in atomic detail while also relaxing the frozen solute assumption and allowing for thermal fluctuations of the solute \cite{chen2014validating, chatzimagas2022predicting}. The relaxation of the frozen solute approximation is particularly relevant for IDPs since they assume many inter-converting conformations in solution. With these advances in computational scattering methods, the trajectories from an explicit solvent simulation can be translated to scattering intensities in great detail for a one-to-one comparison with experimental data. 

In this work, we isolate the role of the protein and water force fields in achieving a generalizable simulation method by using the SAXS data collected for three sequences of EK polyampholytes. We have also uploaded a new, open-source scattering model called `Small and Wide Angle X-ray Scattering for All Molecular Dynamics Engines' (SWAXS-AMDE) to \href{https://github.com/rohansadhikari96/SWAXS-AMDE}{GitHub}. SWAXS-AMDE calculates the background subtracted scattering profiles of biomolecular systems by accounting for both the solute and solvent contributions in atomic detail. It also incorporates the effect of a thermally fluctuating solute on the scattering profile. SWAXS-AMDE is not housed in any simulation engine and is capable of handling binary trajectory files and topology files from all of the popular MD engines [AMBER \cite{case2005amber}, GROMACS \cite{berendsen1995gromacs}, NAMD \cite{phillips2020scalable}, and OpenMM \cite{eastman2017openmm}]. Our results using SWAXS-AMDE indicate that the choice of a protein force field is much more significant than previously thought in achieving a generalized simulation model. Our results also show the advantages of a detailed scattering analysis (without any free-parameters) in validating force field modifications.

The rest of this paper is structured as follows. The synthesis and SAXS characterization of EK polyampholytes, the rationale for choosing the simulation models used in this work, relevant details about the MD simulations, and the theory of the scattering model used are explained in \textbf{Sec.~\ref{sec:meth}}. In \textbf{section \ref{sec:res_disc}}, the scattering profiles from MD simulations are compared to experiments and the performance of the various simulation models (versus experiments) are understood by analyzing their predicted radius of gyration (R$\mathrm{_g}$) and dihedral angle distributions.

\section{Methods} \label{sec:meth}

\subsection{Synthesis of EK Polyampholytes}

Sequence-precise polyampholyte peptides [(EK)$\mathrm{_{16}}$, (E$\mathrm{_2}$K$\mathrm{_2}$)$\mathrm{_{8}}$, and (E$\mathrm{_4}$K$\mathrm{_4}$)$\mathrm{_{4}}$] were synthesized using Fmoc-based solid phase peptide synthesis (SPPS) with all L-chiral monomers. After synthesis, the samples were purified using reversed-phase HPLC (RP-HPLC), ion exchanged, dialyzed, and lyophilized to ensure monodisperse peptide and to reduce counterion effects. A detailed description of the synthesis and purification method can be found in the SI of Ref. \citenum{shi2023influence}. SAXS experiments were conducted at beamline 16-ID LiX of the National Synchrotron Light Source II (NSLS-II) at the Brookhaven National Laboratory. SAXS sample solutions were prepared at 0.5 wt\%. The low solute concentration reduces the possibility of solute-solute intermolecular interactions. Static scattering experiments were measured using a flow cell set up \cite{yang2020solution} with X-ray energy of 13.5 keV at a detector distance of 3.73 m to 0.35 m (0.007 – 0.5 \AA$^{-1}$) and a temperature of 293.15 K for 0.5 s at least 7 times. The spectra reduction, averaging, and background subtraction were carried out using the Jupyter Notebook provided by Beamline 16-ID LiX. 1D spectra analysis for Guinier analysis was conducted using the BioXTAS RAW \cite{hopkins2017bioxtas} package and Jupyter Notebooks from Ref.~\citenum{shi2023influence}. 

The MD simulations and explicit water scattering calculations detailed in the rest of this manuscript are chosen to perform a direct comparison to the measured SAXS data. Such a direct comparison to experiments provides an opportunity to discern the contributions of the protein and water force fields in accurately modeling both folded proteins and IDPs.    

\subsection{Simulation Design}

 Before explaining our choice of simulation models, we provide a brief review of the literature to motivate the need for isolating the protein and water model contributions towards generalizability.

In an analysis of backbone dihedral angles for the RS peptide (an archetypal IDP), the \\ CHARMM36 protein force field was found to overpopulate the left-handed $\mathrm{\alpha}$-helix amino acid conformations \cite{rauscher2015structural}. Since IDPs frequently undergo $\mathrm{\alpha}$-helix to $\mathrm{\beta}$-sheet transitions in solution, reweighting the dihedral angle correction maps (CMAPs) to accurately represent the amino acid conformations was hypothesized to remove the tendency for generating over-compact IDP ensembles. The resulting CHARMM36m force field (with TIP3P waters) produced extended conformations for the RS peptide but was found to produce over-compact ensembles for both the HIV-1 integrase domain (IN) and the cold-shock protein (CspTm) \cite{huang2017charmm36m}. These developments have led to the belief that the water model plays the decisive role in accurately predicting macromolecular conformations of IDPs. However, detailed hydration free energy calculations for archetypal polypeptides have shown that the protein intramolecular interactions play a key role in determining the macromolecular conformations adopted by biomolecules in solution \cite{asthagiri2017intramolecular, tomar2016importance}. A large part of this dilemma in assessing the respective roles of the protein and water models in obtaining accurate biomolecular conformations arises due to the difficulty in disentangling their individual contributions in explicit water simulations.

\begin{table}
    \centering
    \renewcommand{\arraystretch}{2.0}
        \begin{tabular}{ >{\centering\arraybackslash}m{3.0cm} >{\centering\arraybackslash}m{3.0cm} >{\centering\arraybackslash}m{1.5cm} >{\centering\arraybackslash}m{2.2cm} >{\centering\arraybackslash}m{4.2cm}}
            \hline\hline
            Force field combination label & Protein force field & CMAPs &  Water model &  Water-Water $\mathrm{\epsilon_{LJ}}$ (J/mol)\\
            \hline
                TFF99 & AMBER ff99SB (A99) \cite{hornak2006comparison} & No & TIP3P \cite{jorgensen1983comparison} & 636.4 \cite{izadi2014building} \\
                OFF99 & AMBER ff99SB (A99) \cite{hornak2006comparison} & No & OPC \cite{izadi2014building} &  890.36 \cite{izadi2014building} \\
                FF19O & AMBER ff19SB (A19) \cite{tian2019ff19sb} & Yes & OPC \cite{izadi2014building} &  890.36 \cite{izadi2014building} \\  
            \hline\hline
        \end{tabular}
        \caption{Relevant force field parameters for the three force field combinations used in this study.}
        \label{tab:ff_para}
\end{table}

In order to isolate the specific roles of the protein force field and the water model in obtaining a generalizable simulation method, we employ the following simulation scheme. The combination of the AMBER ff99SB (A99) protein force field and the TIP3P water model (TFF99) is used as a benchmark to assess the changes due to the recent improvements. TFF99 was tuned for folded proteins and is known to predict overly compact conformations for IDPs \cite{shabane2019general, rauscher2015structural, piana2015water}. A combination of the A99 model and OPC waters (OFF99) is used to understand the relative changes due to the inclusion of OPC waters. Shabane et al.~proposed the OFF99 combination as a generalized model \cite{shabane2019general}. The increased water-water dispersion interactions in the OPC waters (in comparison to TIP3P waters), also increases the polypeptide-water hydrophilic interactions (in accordance with the Lennard-Jones mixing rule) which can lead to extended polypeptide ensembles. Comparison of R$_\mathrm{g}$ predictions to experiments showed the OPC water model to remove much of the over-compaction of ensembles reported with the TIP3P model (for representative IDPs) \cite{shabane2019general}. Finally, we combine the recently tuned AMBER ff19SB (A19) protein force field with the OPC waters (FF19O) to understand the effect of the protein force field on generalizability. The major differences between the different force field combinations that are relevant to this work are shown in \textbf{Tab.~\ref{tab:ff_para}}. 

\vspace{0.35cm}

\begin{figure*}[ht!]
    \begin{center}
        \includegraphics[width=16.5cm]{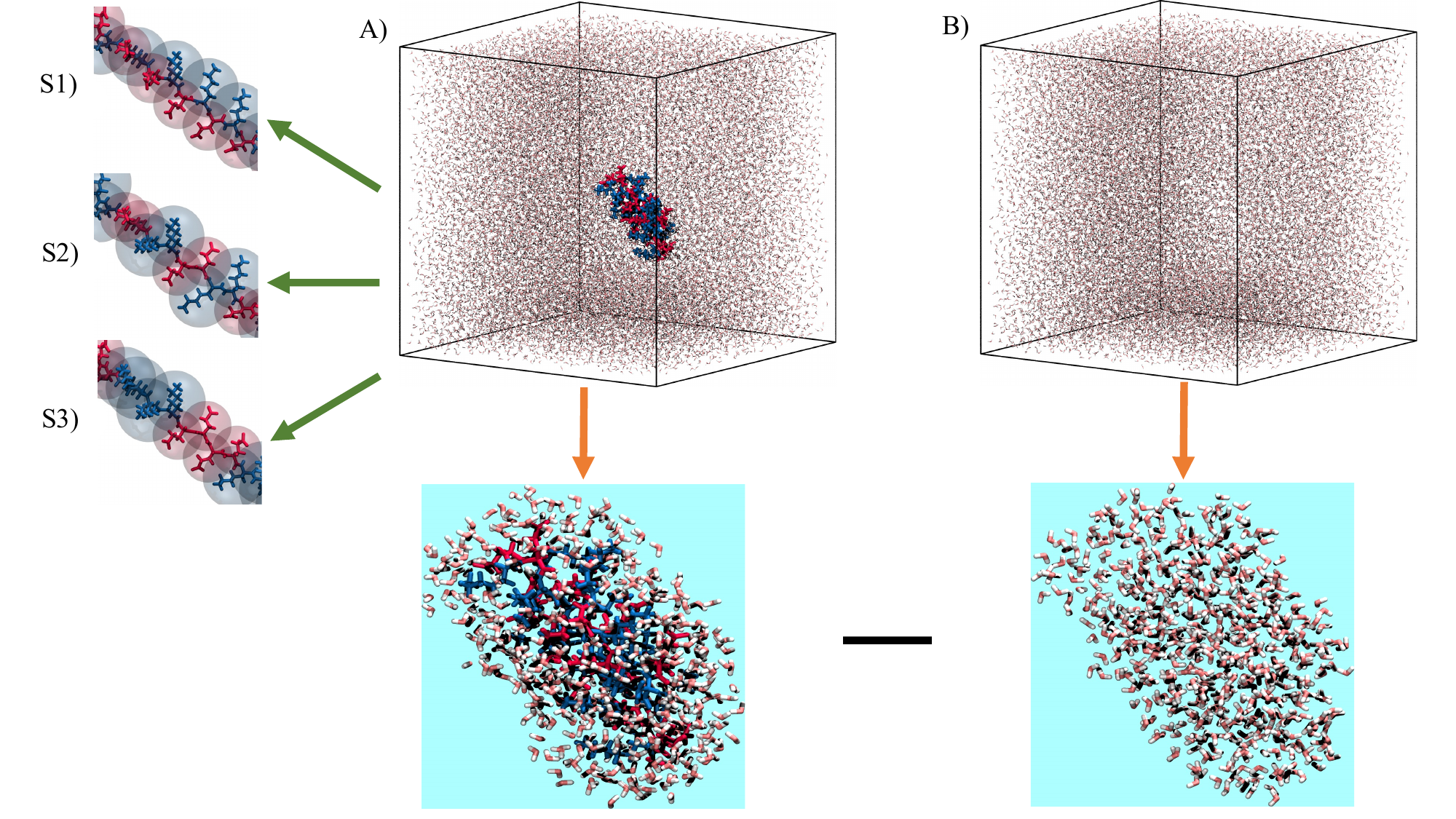}
        \caption{\textbf{Graphical representation of the explicit water simulation and the explicit water scattering calculation for three sequences of EK polyampholytes.} The three polyampholyte sequences (EK)$_{16}$, (E$_2$K$_2$)$_8$, and (E$_4$K$_4$)$_4$ are shown in (S1), (S2), and (S3) respectively. The explicit water simulation of the polyampholyte and the process of carving out a volume of the simulation box for the scattering calculation are shown in (A). The bulk water simulation used for background subtraction of the scattering intensities is depicted in (B).}
        \label{fig:methods}
    \end{center}
\end{figure*}

\subsection{Simulation Details}

The TFF99, OFF99, and FF19O combinations are used to simulate three sequences of EK polyampholytes [(EK)$_{16}$, (E$_2$K$_2$)$_8$, and (E$_4$K$_4$)$_4$] each in explicit water (nine simulations in total). The explicit water simulations for the three different sequences of EK polyampholytes are graphically depicted in \textbf{Fig.~\ref{fig:methods}}. All nine simulations contain a single polyampholyte chain in solution. All simulations are performed using the AMBER20 \cite{case2021amber} simulation package. The OFF99 and FF19O simulations are performed in the NPT (293.15 K and 1 atm pressure) ensemble with 57600 waters which gives a cubic box of size $\approx$ 120*120*120 \AA, while the TFF99 simulations are performed in the NPT ensemble (293.15 K and 1 atm pressure) with 29800 waters which gives a cubic box of size $\approx$ 96*96*96 \AA. The minimum distance between any solute atom and the simulation wall is always greater than 10 \AA \, in all directions. This nullifies the possibility of periodicity artifacts. In addition, the explicit water scattering method requires at least a 7 \AA \, hydration shell around the solute \cite{park2009simulated}, hence, our simulation boxes are larger than those used in other force field validation studies \cite{shabane2019general, rauscher2015structural, huang2017charmm36m}. The OFF99 and FF19O simulation box sizes were also set to 96*96*96 \AA \, initially, but the polyampholytes assumed extended conformations that were too close to the wall. Therefore, the box sizes were increased to 120*120*120 \AA \, and the simulations were repeated. TFF99 simulations did not have this issue with a 96*96*96 \AA \, simulation box. 

Simulations of IDPs in water are known to converge only after multiple $\mu$s. Hence, Shabane et al. propose the use of a 4 fs time step with hydrogen mass repartitioning (HMR) for the solute. We follow the same procedure and use a 4 fs time step in our simulations by increasing the hydrogen masses in the solute to 3.024 atomic mass units. In HMR, the mass of hydrogen's parent atom is decreased to ensure an unchanged solute mass \cite{hopkins2015long}. All nine simulations are run for 8 $\mu$s each, the polyampholyte and water coordinates in the second half of the simulation are written to a trajectory file once every 40 ps to obtain 10$\mathrm{^5}$ simulation frames for subsequent analysis.

Testing the generalizability of simulation models requires a more detailed validation measure than the mean R$_\mathrm{g}$. SAXS provides information about the biomolecular system across multiple length scales, hence we use SAXS profiles as a validation measure for the conformational ensembles generated by force fields. EK polyampholytes have been studied using implicit solvent and coarse grained models \cite{das2013conformations, devarajan2022effect}. In addition to the limitations of implicit solvent models discussed in the literature \cite{weber2012regularizing, adhikari2022hydration}, the implicit solvent approach does not consider density variations in the solvent layer immediately surrounding the solute, making a detailed scattering calculation (that accounts for all density variations relative to the bulk solvent) impossible.

\subsection{The SWAXS-AMDE Model} \label{sec:methods-swaxs-amde}

In order to directly compare the scattering profiles from simulations to experiments, the development of the SWAXS-AMDE model begins with the equation shown below. 

\begin{eqnarray}
     \Delta I(q) = I_{\mathrm{A}}(q) - I_{\mathrm{B}}(q),
\label{eq:back_sub}
\end{eqnarray}

In Eq.~\ref{eq:back_sub}, $\Delta I(q)$ represents the background subtracted scattering intensities, $I_{\mathrm{A}}(q)$ represents the scattering intensities from the biomolecule in solvent simulation, and $I_{\mathrm{B}}(q)$ represents the scattering intensities from the bulk solvent (buffer) simulation. The simulated background subtraction procedure for explicit solvent simulations is graphically represented in \textbf{Fig.~\ref{fig:methods}}.

The form of Eq.~\ref{eq:back_sub} most relevant to explicit solvent simulations was derived by Park et al.~\cite{park2009simulated} and is shown in Eq.~\ref{eq:expl_sca}.  Eq.~\ref{eq:expl_sca} allows for the translation of the atomic coordinates stored in trajectory files to background subtracted scattering intensities. $A(\textbf{q})$ in Eq.~\ref{eq:expl_sca}, represents the complex scattering amplitude calculated within a control volume from the biomolecule in solvent simulation. $B(\textbf{q})$ in Eq.~\ref{eq:expl_sca}, represents the complex scattering amplitudes calculated within the same control volume for the bulk solvent simulation. $\langle ... \rangle'$, $\langle ... \rangle''$, and $\langle ... \rangle_{\mathrm{\Omega}}$ in Eq.~\ref{eq:expl_sca} represent an ensemble average over all the simulation frames in the biomolecule in solvent simulation, an ensemble average over all the simulation frames in the bulk solvent simulation, and an orientational average of the scattering intensities respectively. For Eq.~\ref{eq:expl_sca} to be valid, the control volume needs to be large enough to accommodate all non-bulk like behavior of the solvent. That is, the control volume should encapsulate all regions in space (around the solute) where the electron density of the solvent  differs from the electron density of the bulk solvent. For polypeptides in water, moving a distance 7 \AA \, away from all solute atoms is found to be enough to satisfy this criteria \cite{park2009simulated, knight2015waxsis}. Eq.~\ref{eq:expl_sca} is also the central equation of the SWAXS-AMDE model.       

\begin{eqnarray}
        \Delta I(q) = \bigg \langle \big | \langle A(\textbf{q}) \rangle' - \langle B(\textbf{q}) \rangle'' \big | ^2 + \big[ \langle |A(\textbf{q})|^2 \rangle'  - \big | \langle A(\textbf{q}) \rangle' \big |^2 \big] \big[ \langle |B(\textbf{q})|^2 \rangle''  - \big | \langle B(\textbf{q}) \rangle'' \big |^2 \big] \bigg \rangle_{\mathrm{\Omega}}
        \label{eq:expl_sca}
\end{eqnarray}   

Park et al.~\cite{park2009simulated} assumed a frozen biomolecule in their work, because, it is much easier to define a control volume (which encapsulates all solvent density variations) when there are no thermal fluctuations of the solute. Chen and Hub described the definition of the control volume for simulations in which the biomolecule is not restrained and is allowed to adopt different conformations (implemented in the WAXSiS web server and the GROMACS-SWAXS package) \cite{chen2014validating, knight2015waxsis}. Accounting for thermal fluctuations is crucial for calculating the scattering profiles of IDPs as they adopt an ensemble of inter-converting conformations in solution. SWAXS-AMDE also allows for a thermally fluctuating solute and is hence capable of calculating a detailed scattering profile for both IDPs and folded proteins. Although the central scattering equation in GROMACS-SWAXS and SWAXS-AMDE are the same, GROMACS-SWAXS and the WAXSiS web server require the trajectory files to be in the binary xtc format (used in the GROMACS software). However, SWAXS-AMDE is blind to the choice of the MD simulation software and is capable of reading binary trajectories output by MD engines such as AMBER, GROMACS, NAMD, and OpenMM. This is because SWAXS-AMDE is based on the Python programming language and uses the MDTraj \cite{mcgibbon2015mdtraj} library to read binary trajectory files. Additional details about the SWAXS-AMDE model are provided in the \textbf{supporting information Sec.~S.2.}

Continuum scattering models such as CRYSOL have previously been paired with explicit water simulations for force field validation \cite{rauscher2015structural, huang2017charmm36m, rieloff2021molecular}. In our previous study, however, we found that the presence of free parameters in CRYSOL can cancel out the errors from atomistic force fields (specifically the generation of over-compact IDP conformations) \cite{adhikari2024quantifying}. Hence, we use SWAXS-AMDE to perform a detailed (without free-parameters) scattering analysis for all of the simulated systems in this work. SWAXS-AMDE also offers flexibility for force field validation studies since it can analyze the ensembles generated using recently tuned force fields on a variety of MD simulation packages.

\section{Results and Discussion} \label{sec:res_disc}

\subsection{Comparison of SAXS Profiles} 

\begin{figure*}[ht!]
    \begin{center}
        \includegraphics[width=16.5cm]{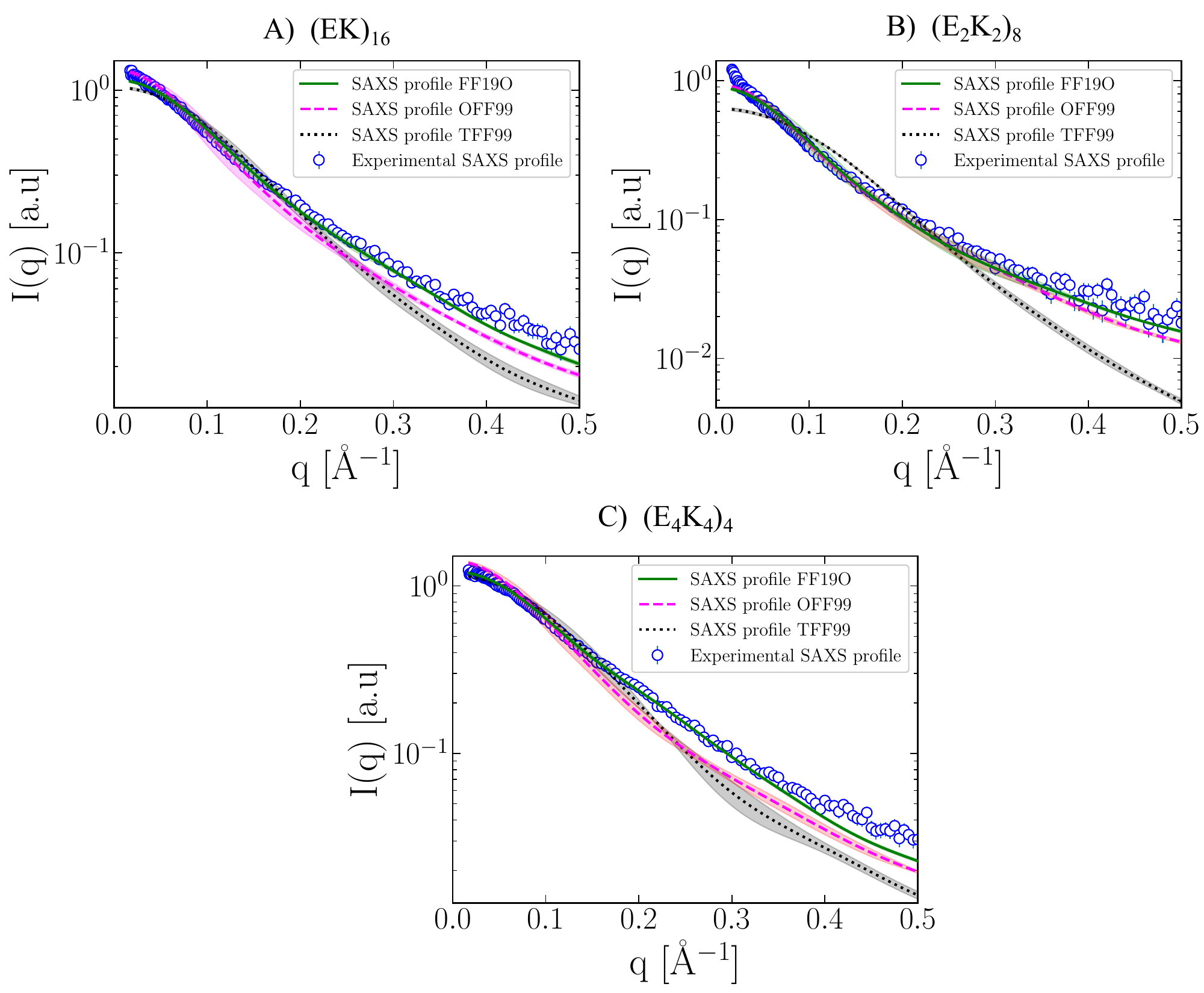}
        \caption{\textbf{Comparison of computed and experimental scattering profiles for three sequences of EK polyampholytes.} (A) Comparison of scattering intensities for the (EK)$\mathrm{_{16}}$ polyampholytes. (B) Comparison of scattering intensities for the (E$\mathrm{_2}$K$\mathrm{_2}$)$\mathrm{_{8}}$ polyampholytes. (C) Comparison of scattering intensities for the (E$\mathrm{_4}$K$\mathrm{_4}$)$\mathrm{_{4}}$ polyampholytes. The computed scattering profiles for the AMBER ff99SB-TIP3P (TFF99) combination are shown in black (dotted lines). The computed scattering profiles for the AMBER ff99SB-OPC (OFF99) combination are shown in magenta (dashed lines). The computed scattering profiles for the AMBER ff19SB-OPC combination (FF19O) are shown in green (solid lines). The shaded regions around the computed scattering profiles represent the error bars from the scattering computation. The experimental scattering profiles are shown as blue symbols (with error bars). The TFF99 data used in this figure is obtained from RSA's thesis (Ref. \citenum{sridhar2024simulation}).}
        \label{fig:scat_res}
    \end{center}
\end{figure*}

Out of the 10$\mathrm{^5}$ simulation frames generated in the last 4 $\mu$s of each of the nine simulations, we select 5000 simulation frames each for analysis using SWAXS-AMDE. The methodology for selecting 5000 simulation frames for analysis is detailed in the \textbf{supporting information Sec.~S.4}. An explicit water scattering calculation for all the frames in the simulation trajectory would be prohibitively expensive. Chen et al.~\cite{chen2014validating} showed that 500 simulation frames are sufficient to get a representative scattering curve from MD simulations. Also, we find that the error bars from our scattering computation in \textbf{Fig.~\ref{fig:scat_res}} are small, hence, we infer that our choice of 5000 simulation frames for analysis is sufficient.

The scattering profiles from simulations are compared to the experimental data for the three polyampholyte sequences [(EK)$\mathrm{_{16}}$, (E$\mathrm{_2}$K$\mathrm{_2}$)$\mathrm{_{8}}$, and (E$\mathrm{_4}$K$\mathrm{_4}$)$\mathrm{_{4}}$] in \textbf{Fig.~\ref{fig:scat_res}(A)-(C)} respectively. The AMBER ff99SB-TIP3P waters combination (TFF99) performs the poorest in comparison to the experimental SAXS data (both visually and in the $\chi$ value reported in the \textbf{supporting information Sec.~S.5}). This is expected, as it is extensively reported in literature that the TFF99 combination produces overly compact conformations for IDPs \cite{shabane2019general, rauscher2015structural, piana2015water}. Replacing the water model in the combination from the TIP3P waters to the OPC waters (to obtain the OFF99 combination) results in noticeable improvements in the scattering profile comparison versus experiments. Hence indicating that the OPC waters are an essential part of a generalized force field combination. This is in good agreement with the conclusions drawn by Shabane et al.~\cite{shabane2019general}. Most notably, replacing the protein force field (in the OFF99 combination) from the A99 protein force field to the A19 protein force field (FF19O) results in the best performing (lowest $\mathrm{\chi}$) scattering profiles versus experiments. To understand these results in more detail, we first look at the distributions for the polyampholyte conformations [by analyzing the probability density distribution of the radius of gyration (R$\mathrm{_g}$)] predicted by these three combinations. Second, we analyze the distribution of the secondary structures predicted by these three combinations using Ramachandran plots \cite{ ramachandran1968conformation}.

\subsection{Distribution of R$\mathrm{_\textbf{g}}$} 

\begin{figure*}%[tbhp]
    \begin{center}
        \includegraphics[width=16.5cm]{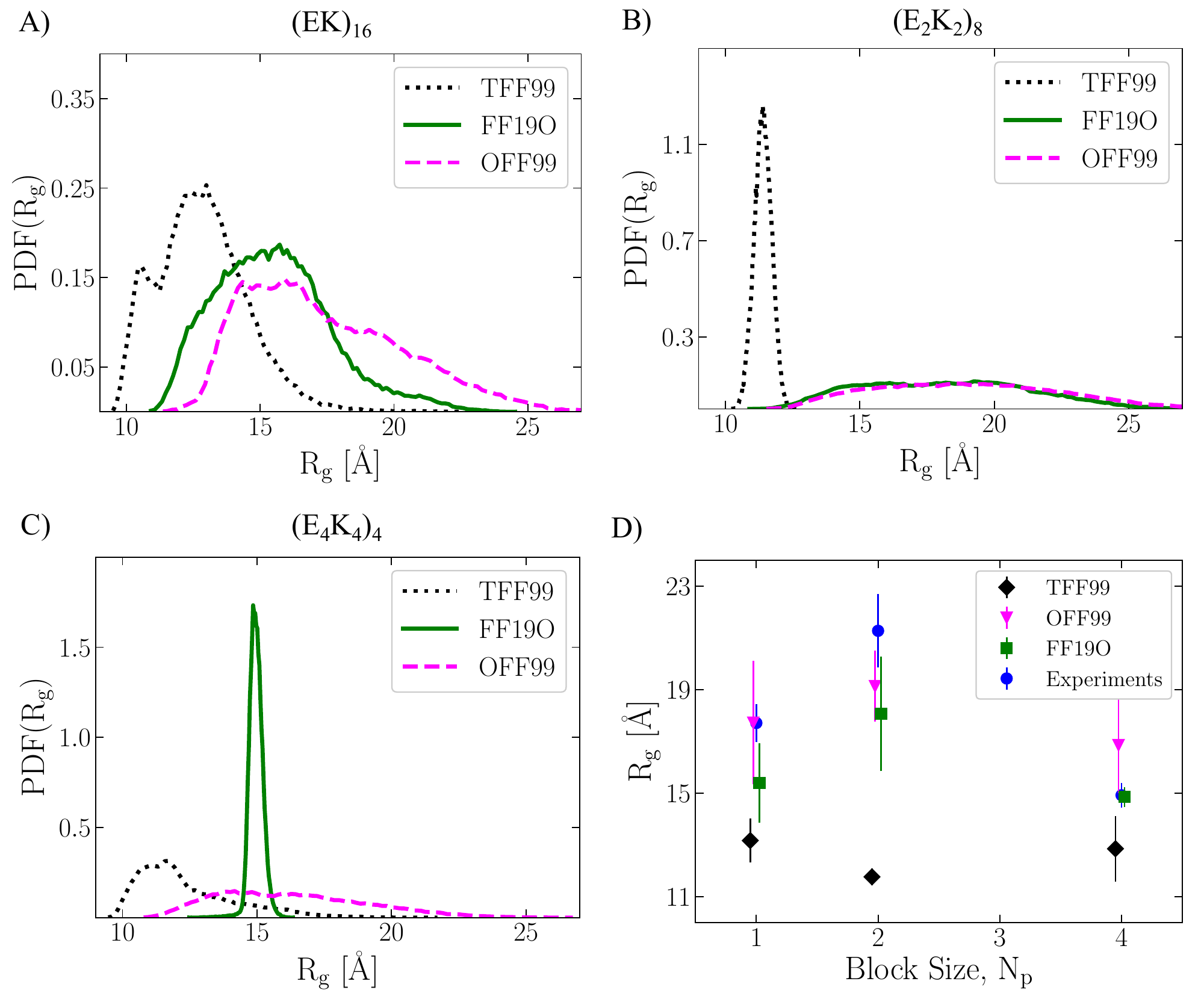}
        \caption{\textbf{Distribution of the radii of gyration (R$\mathrm{_g}$) of polyampholytes from the three simulation methods and comparison of Guinier analysis based R$\mathrm{_g}$ from simulations to experiments.} The R$\mathrm{_g}$ distributions (from the three simulation combinations) for the three polyampholyte sequences [(EK)$\mathrm{_{16}}$, (E$\mathrm{_2}$K$\mathrm{_2}$)$\mathrm{_{8}}$, and (E$\mathrm{_4}$K$\mathrm{_4}$)$\mathrm{_{4}}$] are shown in (A)-(C) respectively. The Guinier analysis based R$\mathrm{_g}$ from the three combinations of force fields are compared to the experimental R$\mathrm{_g}$ in (D). The block size of the polyampholyte sequences for TFF99, OFF99, and FF19O are slightly offset in (D) for better visibility of the error bars. The TFF99 data used in this figure is obtained from RSA's thesis (Ref. \citenum{sridhar2024simulation}).}
        \label{fig:rg_dist}
    \end{center}
\end{figure*}

The distribution of R$\mathrm{_g}$ is a good metric to estimate the degree of disorder predicted by the simulation model \cite{shabane2019general}. A wide R$\mathrm{_g}$ distribution indicates a disordered polyampholyte in solution with an ensemble of inter-converting conformations. A narrow R$\mathrm{_g}$ distribution indicates a more ordered polyampholyte with a stable preferred conformation in solution. The R$\mathrm{_g}$ distribution predicted by simulation methods for the three polyampholyte sequences are shown in \textbf{Fig.~\ref{fig:rg_dist}(A)-(C)} respectively. The R$\mathrm{_g}$ distribution for all three simulation methods are calculated from the polyampholyte conformations collected in the production phase of the simulation (10$\mathrm{^5}$ simulated conformations for each polyampholyte).

OFF99 and FF19O predict the highest degree of disorder (widest R$\mathrm{_g}$ distribution) for the (E$\mathrm{_2}$K$\mathrm{_2}$)$\mathrm{_8}$ polyampholyte sequence. The SAXS profiles for these two models are also similar for (E$\mathrm{_2}$K$\mathrm{_2}$)$\mathrm{_8}$ and compare well with the experimental SAXS profile. The R$\mathrm{_g}$ obtained from a Guinier analysis of the scattering profile for (E$\mathrm{_2}$K$\mathrm{_2}$)$\mathrm{_8}$ is the highest within the respective predictions of both the FF19O and OFF99 models. A trend that is also captured by the experimental data in \textbf{Fig.~\ref{fig:rg_dist}(D)}. The R$\mathrm{_g}$ distribution for (E$\mathrm{_2}$K$\mathrm{_2}$)$\mathrm{_8}$ from FF19O and OFF99 indicate that it exists in an ensemble of inter-converting conformations in solution. FF19O predicts a lower degree of disorder for (EK)$\mathrm{_{16}}$ compared to OFF99, this translates to a better SAXS performance (lower $\mathrm{\chi}$) of FF19O when compared to OFF99. The R$\mathrm{_g}$ distribution for (E$\mathrm{_4}$K$\mathrm{_4}$)$\mathrm{_4}$ predicted by the FF19O model is the most notable. The FF19O model predicts an extremely narrow R$\mathrm{_g}$ distribution indicating a preferred stable conformation in solution for (E$\mathrm{_4}$K$\mathrm{_4}$)$\mathrm{_4}$. For (E$\mathrm{_4}$K$\mathrm{_4}$)$\mathrm{_4}$, the R$\mathrm{_g}$ from Guinier analysis predicted by FF19O is in excellent agreement with experiments. The error bars from Guinier analysis are the lowest for (E$\mathrm{_4}$K$\mathrm{_4}$)$\mathrm{_4}$ from both FF19O and experiments. This indicates that (E$\mathrm{_4}$K$\mathrm{_4}$)$\mathrm{_4}$ in fact has a stable preferred conformation in solution as predicted by FF19O.  

The property of interest for a direct comparison of force field predictions to experiments is highly debated. R$\mathrm{_g}$ is usually used as a metric to validate simulation methods versus experiments, however, in addition to offering only a single data point for comparison, the error in R$\mathrm{_g}$ estimation from Guinier analysis for IDPs is reported to be as high as 10 \% \cite{zheng2018extended}. We believe the scattering intensities [I(q)] are the property of interest to be validated versus experiments, as it offers the possibility of validating simulation results across multiple length scales. In addition, detailed scattering models (without free-parameters) have been developed to predict the I(q) from simulations and make a direct comparison with experiments.

Based on the R$\mathrm{_g}$ distribution predicted by the force fields, OFF99 always predicts disordered conformations for all three sequences. The conformations predicted by TFF99 is overly compact for all three sequences as expected [see \textbf{Fig.~\ref{fig:rg_dist}(D)}]. The R$\mathrm{_g}$ distribution predicted by FF19O includes both disordered (wide distribution of R$\mathrm{_g}$) and ordered (narrow distribution of R$\mathrm{_g}$) conformations. Hence, FF19O shows great promise in the search for a generalizable simulation method. To understand why FF19O is the most sensitive to the polyampholyte sequence, we analyze the amino acid conformations predicted by the three force field combinations using Ramachandran plots.

\subsection{Ramachandran Plots} 

Ramachandran plots \cite{ramachandran1968conformation} are a useful visual tool for categorizing the dihedral angle pairs ($\mathrm{\phi}$-$\mathrm{\psi}$ angles in the polypeptide backbone) predicted by simulations. Broadly speaking, the $\mathrm{\beta}$-sheet conformations can be found on the top left quadrant of a Ramachandran plot, the left-handed $\mathrm{\alpha}$-helix ($\mathrm{\alpha_L}$) conformations can be found on the top right quadrant of a Ramachandran plot, and the right-handed $\mathrm{\alpha}$-helix ($\mathrm{\alpha_R}$) conformations can be found on the bottom left quadrant of a Ramachandran plot. The $\mathrm{\beta}$-sheet conformations are more extended in comparison to the $\mathrm{\alpha}$-helix conformations. Dihedral angle correction maps (CMAPs) were introduced to protein force fields in order to penalize $\mathrm{\phi}$-$\mathrm{\psi}$ angles that have a low probability of occurrence in nature. 

\begin{figure*}%[tbhp]
    \begin{center}
        \includegraphics[width=16.5cm]{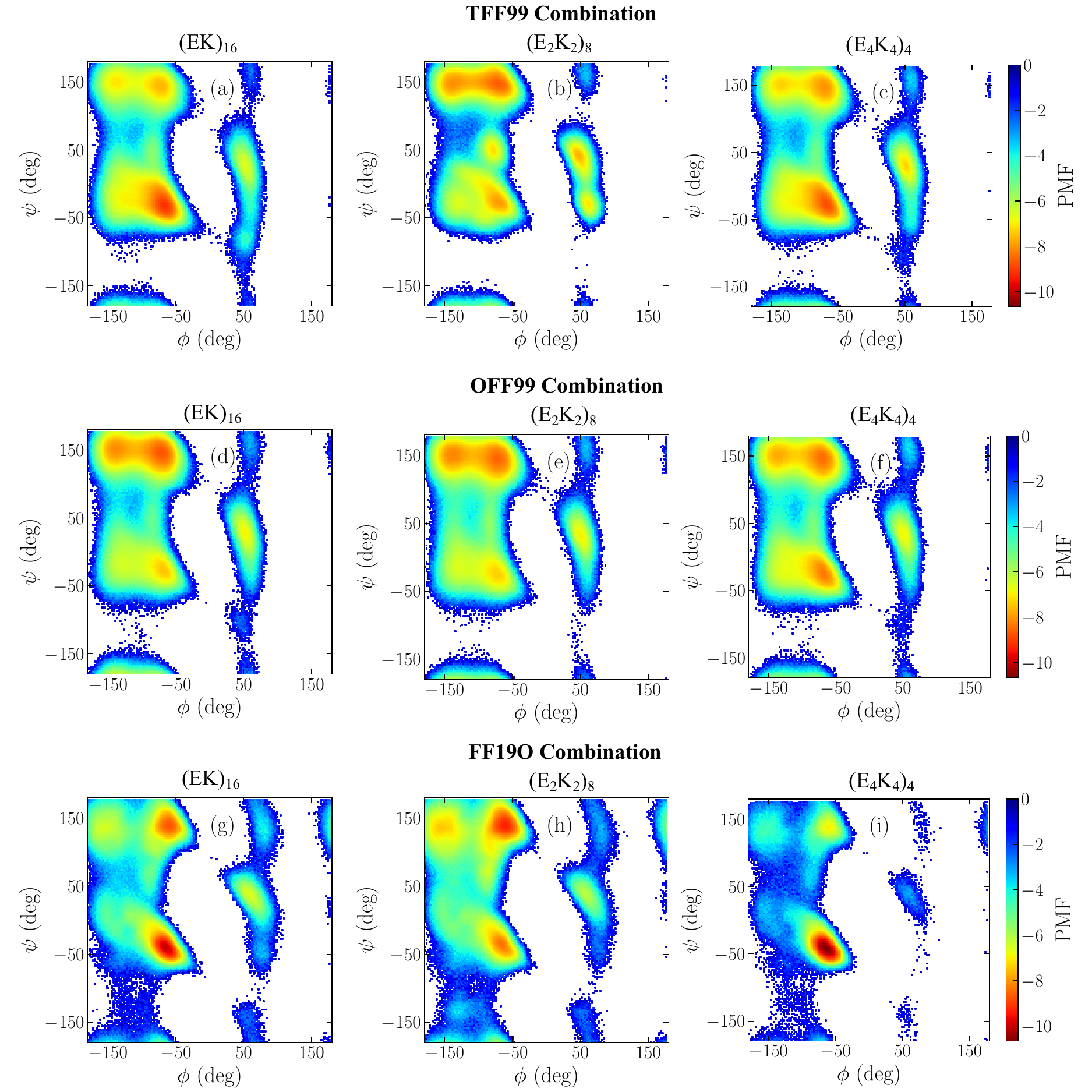}
        \caption{\textbf{Ramachandran plots for all nine simulations (three force field combinations * three polyampholyte sequences).} All of the Ramachandran plots are obtained from the last 4 $\mu$s of the simulation. The Ramachandran plots for the TFF99 combination are shown for the three polyampholyte sequences [(EK)$\mathrm{_{16}}$, (E$\mathrm{_2}$K$\mathrm{_2}$)$\mathrm{_{8}}$, and (E$\mathrm{_4}$K$\mathrm{_4}$)$\mathrm{_{4}}$] in (A)-(C) respectively. The Ramachandran plots for the OFF99 combination are shown for the three polyampholyte sequences [(EK)$\mathrm{_{16}}$, (E$\mathrm{_2}$K$\mathrm{_2}$)$\mathrm{_{8}}$, and (E$\mathrm{_4}$K$\mathrm{_4}$)$\mathrm{_{4}}$] in (D)-(F) respectively. The Ramachandran plots for the FF19O combination are shown for the three polyampholyte [(EK)$\mathrm{_{16}}$, (E$\mathrm{_2}$K$\mathrm{_2}$)$\mathrm{_{8}}$, and (E$\mathrm{_4}$K$\mathrm{_4}$)$\mathrm{_{4}}$] sequences in (G)-(I) respectively. The TFF99 data used in this figure is obtained from RSA's thesis (Ref. \citenum{sridhar2024simulation}).}
        \label{fig:rama_plots}
    \end{center}
\end{figure*}

The polyampholyte sequences used in this study have 32 amino acid residues each. To obtain the Ramachandran plots shown in \textbf{Fig.~\ref{fig:rama_plots}}, the 3.2*10$\mathrm{^6}$ (32 dihedral pairs for each polyampholyte conformation * 10$\mathrm{^5}$ simulation frames) dihedral angle pairs from each simulation are sorted into bins of size 2.5$\mathrm{^0}$*2.5$\mathrm{^0}$. The Ramachandran plots collected in the first half of the simulation (see \textbf{supporting information Sec.~S.6}) characteristically match those shown in \textbf{Fig.~\ref{fig:rama_plots}} for all nine simulations, indicating a convergence of the MD simulations. The Ramachandran plots for the TFF99 [\textbf{Fig.~\ref{fig:rama_plots}(A)-(C)}] and OFF99 [\textbf{Fig.~\ref{fig:rama_plots}(D)-(F)}] combinations are more permissive in comparison to the FF19O combination [\textbf{Fig.~\ref{fig:rama_plots}(G)-(I)}]. This is understandable given that the ff99SB (A99) protein force field does not include CMAP corrections.

The Ramachandran plots for OFF99 show little distinction between the three polyampholyte sequences studied. OFF99 predicts sizable populations of $\mathrm{\beta}$-sheet, $\mathrm{\alpha_L}$, and $\mathrm{\alpha_R}$ conformations for all three sequences, which is closely associated with its wider R$\mathrm{_g}$ distribution [\textbf{Fig.~\ref{fig:rg_dist}(A)-(C)}] for all three polyampholyte sequences. Comparatively, the FF19O Ramachandran plots show more distinction between the three polyampholyte sequences. Between (EK)$\mathrm{_{16}}$ and (E$\mathrm{_2}$K$\mathrm{_2}$)$\mathrm{_8}$, FF19O predicts a higher probability of $\mathrm{\alpha_R}$ for (EK)$\mathrm{_{16}}$ and a higher probability of $\mathrm{\beta}$-sheets for (E$\mathrm{_2}$K$\mathrm{_2}$)$\mathrm{_8}$. This is closely associated with the lower mean R$\mathrm{_g}$ and lower spread in the R$\mathrm{_g}$ distribution for (EK)$\mathrm{_{16}}$ in comparison to (E$\mathrm{_2}$K$\mathrm{_2}$)$\mathrm{_8}$. Notably, FF19O predicts an extremely high population of $\mathrm{\alpha_R}$ and low populations of $\mathrm{\beta}$-sheet and $\mathrm{\alpha_L}$ conformations for (E$\mathrm{_4}$K$\mathrm{_4}$)$\mathrm{_4}$. The predominance of a specific category of amino acid conformations for (E$\mathrm{_4}$K$\mathrm{_4}$)$\mathrm{_4}$ is closely related to FF19O's prediction of a narrow R$\mathrm{_g}$ distribution in \textbf{Fig.~\ref{fig:rg_dist}(C)}, which indicates a preferred macromolecular conformation in solution.        
The R$\mathrm{_g}$ distribution predicted by FF19O is more sensitive to the polyampholyte sequence in comparison to OFF99. We believe FF19O's ability to predict distinct macromolecular ensembles for the three polyampholyte sequences is closely linked to its discernibly distinct dihedral angle distributions for the three sequences. 

The role of the water model is assumed to be central in accurately predicting the macromolecular ensembles of both folded proteins and IDPs. However, we find that the protein force field also plays a key role in obtaining accurate macromolecular ensembles from MD simulations. Introducing/reweighting CMAPs to remove unphysical backbone dihedral angles and/or to retune the relative populations in the dihedral angle space, will not just influence the secondary structures predicted by the force field. Varying the amino acid conformations using CMAPs can usher the polypeptide into entirely different macromolecular conformations. Hence, we conclude that the role of the protein and water models cannot be separated into length scales. The effect of the protein and water model on the generated ensembles are much more intertwined.     

The FF19O SAXS profiles agree with experiments up to $q \, \sim$ 0.3 \AA$\mathrm{^{-1}}$ (length scale of $\sim$ 20 \AA), which is of the order of the macromolecular conformations of the polyampholytes studied here. There is a slight characteristic disagreement between FF19O and experiments in the $q$-range of 0.3-0.5 \AA$\mathrm{^{-1}}$. One possible reason for this disagreement could be the difference in the relative proportions of secondary structures between FF19O and the synthesized polyampholytes. In this study, the secondary structure predictions from simulations are only used to analyze the link between the dihedral angle distributions and the macromolecular ensembles predicted by the force field. In the future, we plan to test the secondary structure predictions made by FF19O using the nuclear magnetic resonance (NMR), circular dichroism (CD), and Fourier transform infrared spectroscopy (FTIR) methods. Next, we list the other possible sources of discrepancy (for FF19O at $q \, >$ 0.3 \AA$\mathrm{^{-1}}$) in the simulation methodology and SAXS data collection.

\subsection{Source of Residual Discrepancies}

(1) Uncertainties in the background subtraction: The computed SAXS profiles in \textbf{Fig.~\ref{fig:scat_res}} are only scaled by a constant factor (f) when minimizing the errors with respect to the experimental data. This is represented in \textbf{Eq.~\ref{eq:chi2}}. 

\begin{eqnarray}
     \chi^2 = \frac{1}{N} \sum_{i=1}^{N} \bigg[ \frac{fI_{comp}(q_i) - I_{exp}(q_i)}{\sigma_{exp}(q_i)} \bigg]^2
\label{eq:chi2}
\end{eqnarray}

A constant parameter is also frequently added to the computed scattering profiles to account for the presence of dark currents during the experimental background subtraction procedure \cite{knight2015waxsis}. This method of error minimization is detailed in the \textbf{supporting information Sec.~S.7}, and it improves the agreement between the computed scattering profiles and the experimental data for all three polyampholyte sequences. 

(2) The effect of the simulation time step: A 4 fs time step with hydrogen mass repartitioning (HMR) is used in this study to access long simulation time scales (8 $\mu$s). Asthagiri et al. found that increasing the simulation time step to 4 fs introduces a noticeable error in the densities predicted by MD simulations \cite{asthagiri2023md, asthagiri2025consequences}. In order to remove the effect of bulk water density fluctuations arising from the finite box size and the finite number of simulation frames, we use a solvent density correction procedure while computing the scattering profiles \cite{chen2014validating}. This should also remove the effects of bulk density variations (due to the increased time step) on the computed scattering profiles. However, if the increased time step causes a characteristically different behavior of waters in the hydration layer, it can influence the computed scattering profiles.

(3) Hydration free energies of the folding transition for OPC waters:
Shabane et al. reported the $\mathrm{\Delta G}$ values for the folding transition of the CLN025 protein using the TIP3P and OPC waters \cite{shabane2019general}. While the OPC waters were found to reduce the over-stabilization of folded structures in comparison to the TIP3P model, it still over-stabilized folded conformations of the protein in comparison to experiments. The residual tendency to over-stabilize folded protein conformations could also explain the discrepancies between the scattering profiles of FF19O and experiments.

\section{Conclusions}

In summary, three sequences of EK polyampholytes were synthesized using solid phase peptide synthesis as described in Ref.~\citenum{shi2023influence}. The polyampholytes with a precise sequence of charged monomers were subsequently subjected to SAXS characterization. Three combinations of protein and water force fields were used to perform atomistic simulations of the three EK polyampholytes in water. The SWAXS-AMDE package was used to perform scattering computations from the simulation trajectory files. Validation of simulated scattering profiles revealed that a combination of the AMBER ff19SB force field for proteins and the OPC water model (FF19O) performed the best versus the experimental SAXS data. The design of our simulations allowed us to quantify the specific improvements due to both the dihedral angle corrections to the protein force field and the OPC water model. A study of the conformational distributions for the three sequences of EK polyampholytes revealed the generalized nature of the FF19O combination. An analysis of the secondary structure of EK polyampholyte sequences using Ramachandran plots revealed that FF19O was the most sensitive to the polyampholyte sequence. Indicating a close link between the diverse secondary structure predictions from FF19O for the three polyampholyte sequences and its generalizable nature. Overall, our results show that the protein force field also plays a central role in obtaining accurate macromolecular ensembles from MD simulations.

We also stress the importance of using a detailed (without free-parameters) scattering computation to accurately extract the information from atomistic MD simulations and validate their predictions. Accounting for the thermal fluctuations of the protein during the scattering computation is crucial for IDPs that adopt an ensemble of inter-converting conformations in solution. In order to facilitate an atomistic scattering computation of a thermally fluctuating polypeptide, we have uploaded the open-source SWAXS-AMDE package to \href{https://github.com/rohansadhikari96/SWAXS-AMDE}{GitHub}. SWAXS-AMDE can read the binary trajectory files from all of the popular MD engines.

\begin{acknowledgement}

RSA and WGC acknowledge Dr.~Dilipkumar N. Asthagiri at the Oak Ridge National Laboratory for his insightful comments on simulation time steps.  RSA acknowledges Dr.~Arjun Valiya Parambathu for sharing his valuable advice on graphics. We gratefully acknowledge the Robert A. Welch foundation (Grant C-1241) for their financial support. ABM thanks the Robert A. Welch foundation for financial support (C-2003-20190330). ABM gratefully acknowledges the National Science Foundation (NSF) CAREER Award (2338550) in the Division of Materials Research (DMR) for financial support. This research used the resources provided by the San Diego Supercomputing Center (SDSC) through the Expanse cluster under a  National Science Foundation (NSF) Advanced Cyberinfrastructure Co-ordination Ecosystem (ACCESS) program (Grant no.~BIO230123). This research used resources of the National Synchrotron Light Source II, a U.S.~Department of Energy (DOE) Office of Science User Facility operated for the DOE Office of Science by Brookhaven National Laboratory under Contract No.~DE-SC0012704. Parts of this work are adapted from the PhD thesis of RSA at Rice University, Houston, TX, USA, titled `Simulation Studies on Polyampholytes can Inform Models of Protein Solution Thermodynamics'

\end{acknowledgement}

\begin{suppinfo}

(1) Additional simulation details. (2) Additional details of the scattering computation. (3) Validation of the SWAXS-AMDE scattering model. (4) Selecting representative simulation frames for scattering analysis. (5) Quantifying errors ($\chi$) from the computed scattering profiles. (6) Ramachandran plots for the polyampholytes during the equilibration phase of the simulation. (7) Quantifying errors in the computed scattering profiles using a constant parameter to account for uncertainties in the background subtraction.

\end{suppinfo}

\begin{codeavailability}

The SWAXS-AMDE model to obtain the background subtracted scattering intensities using the simulation trajectory files is uploaded to Github (\href{https://github.com/rohansadhikari96/SWAXS-AMDE}{https://github.com/rohansadhikari96/SWAXS-AMDE}), and is available for use freely. The data required to run an example calculation with SWAXS-AMDE is provided on  \href{https://zenodo.org/records/15072930?token=eyJhbGciOiJIUzUxMiJ9.eyJpZCI6Ijk5ZDhlYzFmLWQ0MWYtNDEzMC05NDY1LTBlNDYxYWNhMzYyYyIsImRhdGEiOnt9LCJyYW5kb20iOiJiNDYyMDE4ZjQxZTQ3ZTBiNjE5ZmJkMjk1Y2MwY2ZjOCJ9.XszW2ii4-fQDrXF-wsp9doZfgGnypSlnS_ne6kNXNIG7qXqgrkeg24D5Zp_xt4ymYQjWLVO-HppAbJDfRUsP7g}{Zenodo}. The trajectory files for all nine simulations will be made available by the authors upon reasonable request.

\end{codeavailability}

\bibliography{myref}

\includepdf[pages=-]{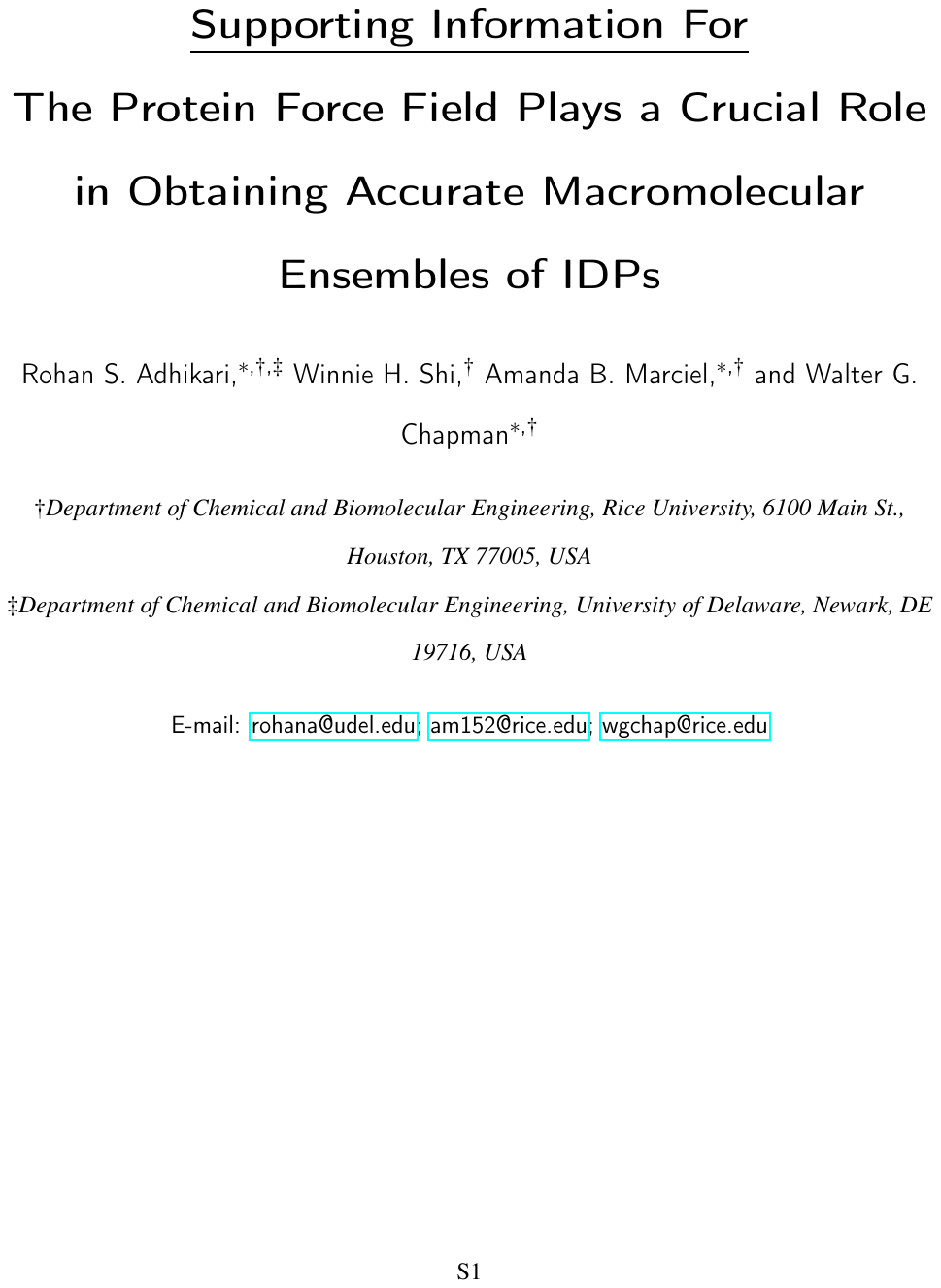} 

\end{document}